\documentclass[a4paper,11pt]{article}
\pdfoutput=1 

\usepackage{jcappub} 

\usepackage[T1]{fontenc} 

 \title{Dilaton generation in propagation of magnetic dipole waves of pulsar in a galactic magnetic field}

\author{M.O.Astashenkov}
\author{A.V.Bedda} 
\author{K.V.Parfenov}
\author{P.A.Vshivtseva} 
\affiliation{ Physics Department, Moscow State University, Moscow 119991 Russia}

\abstract{
	
This study is devoted to dilaton generation during the propagation of magnetic dipole waves from a pulsar in the galactic magnetic field. Dilaton generation occurs at cosmological scales on the order of the coherence lengths of the galactic magnetic field $L_{\mathrm{coh}}$, approximately 100 pc. 
	
The exact solutions of the dilaton field equation in a vacuum and in the interstellar medium with reflective index $n = 1 + \chi$, $\chi \ll 1$ have been obtained, and the angular distribution of emitted dilaton energy has been determined in both cases. It has been shown that the dilaton generation intensity at first increases proportionally to $r^2$, where $r$ is distance from the neutron star to the observation point, then has oscillating behavior. 
The obtained results are applicable only for $r < L_{\mathrm{coh}}$.

	For a millisecond pulsar with a magnetic field $B_S \sim 10^9$ Gauss, located at a distance from Earth on the order of the coherence length of the galactic magnetic field $L_{\mathrm{coh}} \sim 100$ pc, the obtained intensity of the dilaton generation can be greater in 100 times than the analogous intensity produced by rotating magnetic dipole momentum of a pulsar.

 	Based on estimated values, the impact of pulsar and magnetar gravitational fields on magnetic dipole wave radiation is minimal and changes the amplitude of the waves by around 1 percent. For this reason, the effect of the gravitational field on the dilaton formation process can be disregarded in this study.

}

\begin{document}%


\maketitle 
\flushbottom

\section{Introduction}

An additional option within the Standard Model could be the inclusion of a hypothetical particle, called dilaton. In the field theory, dilaton is a scalar field that arises when conformal symmetry is spontaneously broken \cite{nat_light_dil, light_dilaton}, as well as in gravity theories with extra dimensions \cite{Kaluza-klein, string-th, RS-holography}.

Consideration of the dilaton field helps shed light on specific cosmological problems. The existence of the cosmological constant and inflation can be explained through dilaton field theory \cite{dil_and_lambda, dilaton-inflation, dilaton-inflation-2}. Furthermore, exploring black holes with a dilaton may reveal new understanding of black hole thermodynamics \cite{dilaton-black-hole1, dilaton-black-hole2}.
Dilaton also appears in composite Higgs models. In this scenario, a dilaton could serve as a connection between the Standard Model and dark matter particles \cite{dil-portal1, dil-portal2}.

Dilaton is  also predicted in multidimensional theories of gravity, such as the 5-dimensional Kaluza-Klein theory \cite{Kaluza-klein}, which unifies gravitational and electromagnetic interactions, and the superstring theory \cite{string-th}. In these theories, a dilaton is considered as a scalar field that is localized in the diagonal elements of a multidimensional metric tensor $g_{AB}$ corresponding to the coordinates of extra spacetime dimensions. In a 5-dimensional scenario, the dilaton field may be located in metric tensor element $g_{44}$, while elements $g_{00}$, $g_{11}$, $g_{22}$, $g_{33}$ represent standard 4-dimensional space-time \cite{Kaluza-klein}.

In this study, the dilaton is considered in the context of Maxwell-dilaton theory \cite {EMD}, which in particular follows from the multidimensional theories of gravity \cite {Kaluza-klein, string-th} mentioned above. In section \ref{sec:Maxwell-dil}, it will be demonstrated that dilatons can be generated by electromagnetic fields within the context of this theory.

The search for dilatons is actively conducted at the Large Hadron Collider \cite{dil_LHC}, as well as in other experiments \cite{env-dep-dil}. As dilaton can be generated by electromagnetic fields, it can be detected in experiments searching for axion-like particles \cite{ALP-exp-search, NA64}.  Nowadays, only restrictions on their coupling constant \cite{ALP-exp-search, NA64} have been obtained.
 In particular, a restriction has been obtained on the coupling constant of axion-like particles born in the core of the Sun \cite{CAST}.

There is potential to explore dilatons \cite{dil_q_plane_wave, dil_from_pulsar} generated by various configurations of electromagnetic fields in certain astrophysical objects. In particular, it is notable to explore rotating neutron stars (pulsars and magnetars) as sources of dilatons. These objects are known to possess powerful magnetic fields ranging from $10^8$ to $10^{15}$ Gauss \cite{pulsar-catalog}, and emit coherent electromagnetic radiation.
The generation of dilatons by rotating neutron stars is discussed in detail in \cite{dil_from_pulsar}.
Pulsars are also promising sources of axin-like particles \cite{arion_from_pulsar, arion_magn_field, axion-from-pulsar}.

This study will focus on the dilaton generation during the propagation of magnetic
dipole radiation in a galactic magnetic field.
The galactic magnetic field can be considered constant and uniform \cite{gltc_mag_field} on
quite large distances $L_{\mathrm {coh}}$, where $L_{\mathrm {coh}}$ is the coherence length of
magnetic field. For galactic magnetic field $L_{\mathrm {coh}}\approx $100 pc \cite{gltc_legth_coh}.
For this reason, despite galactic magnetic fields being relatively weak $\sim 10^{-5}$ Gauss \cite{gltc_mag_field}, they can still play a substantial role in dilaton generation.

The structure of this paper is as follows. In the section \ref{sec:Maxwell-dil}, the Maxwell-dilaton theory is described, and it is shown that in this theory dilatons can be generated by electromagnetic fields. Then in Section \ref{sec:eq} we introduce the magnetic field of rotating magnetic dipole momentum of a neutron star and a galactic magnetic field. After that, the equation of dilaton generated by the considered configuration of electromagnetic fields is obtained, and in the next section \ref{sec:solution}, its exact solution is found. Next, in the section \ref{sec:I} intensity of dilaton generation is obtained based on the found exact solution. It is shown that the obtained intensity grows as the distance increases. For this reason, in Section \ref{sec:I_n>1} we will take into account a reflective index of the interstellar medium in which magnetic dipole radiation of a rotating neutron star propagates. In section \ref{sec:I_n>1} there is the discussion of dilaton generation solution and intensity taking into account this specific property. Finally, section \ref{sec:conclusion} contains the main conclusion and results.

\section{Maxwell-dilaton theory} \label{sec:Maxwell-dil}

Following the study \cite{dil_from_pulsar}, the density of the Lagrange function for the dilaton interacting with an electromagnetic field can be expressed in the form:
\begin{equation} \label{L}
	{\cal L} =a_0(\partial\Psi)^2+a_1e^{-2{\cal K}\Psi}F_{nm}F^{nm}\;,
\end{equation}
where $a_0,\ a_1$ and $\cal K$ are gauge constants and  $F_{nm}$ is Maxwell tensor. Dilaton is considered as a massless scalar field $\Psi$.

The value of the constant ${\cal K} = 1$ is predicted by superstring theory. The 5-th dimentional Kaluza-Klein theory result gives the value ${\cal K} = \sqrt{3}$. 
It should be noted that in theoretical physics there are other theories that predict the existence of dilaton, but they don't have any commonality with gravity \cite{light_dilaton}.
In this study, constant ${\cal K}$ is believed to be arbitrary. 

The field equations obtained from the density of the Lagrange function \eqref{L} have the form:

\begin{equation} \label{field_eq1}
	\partial_k\partial^k \Psi = - \frac{a_1{\cal K}}{a_0}e^{-2{\cal K}\Psi}F_{nm}F^{nm}\;
\end{equation}
\begin{equation} \label{field_eq2}
	\partial_n \big[e^{-2{\cal K}\Psi} F^{nm}\big] = 0
\end{equation}
Equation \eqref{field_eq1} represents the dilaton field $\Psi$, which exhibits self-interaction. Equation \eqref{field_eq2} represents an equation for the electromagnetic field. It also describes how the dilaton field impacts the electromagnetic.

It is expected that dilaton field is weak, and one can put: $|{\cal K}\Psi|<<1.$ 
Under this approximation, the field equations \eqref{field_eq1}, \eqref{field_eq2} in the Minkowski space-time will take the form:
\begin{equation} \label{psi_equation_<<1}
	(\Delta - {1\over c^2}{\partial^2 \over \partial t^2}) \Psi=\frac{a_1{\cal K}}{a_0}F_{nm}F^{nm}\;
\end{equation}

	\begin{equation} \label{F_mn_eq_<<1}
		\partial_nF^{nm} = 0\;.
	\end{equation} 
The equation \eqref {F_mn_eq_<<1} characterizes Maxwellian electrodynamics in the absence of charges or currents. Therefore, in this approximation, there is no transformation of dilatons into electromagnetic fields.

The invariant $F_{mn}F^{mn}$ is represented in terms of the electric field ${\bf E}$ and magnetic field ${\bf B}$ in the following manner:
\begin{equation} \label{F2_inv}
	F_{nm}F^{nm} = 2({\bf B}^2 - {\bf E}^2)
\end{equation}
In this instance, the equation \eqref{psi_equation_<<1} will be expressed as:
\begin{equation} \label{psi_eq_B2_E2}
	(\Delta - {1\over c^2}{\partial^2 \over \partial t^2}) \Psi={2a_1{\cal K}\over a_0} 
	\big[{\bf B}^2-{\bf E}^2\big]\;,
\end{equation}

According to the equations  \eqref{psi_equation_<<1}, \eqref{psi_eq_B2_E2} in the considered approximation, the source of the dilaton field can be only electromagnetic fields for which the invariant \eqref{F2_inv} is not equal to 0. Thus, the dilaton field can be generated either in the near zone of some electromagnetic source or during the propagation of electromagnetic waves in an external electromagnetic field.

Assume that the invariant of electromagnetic field \eqref{F2_inv} varies periodically with a frequency $\omega$ over time $t$, and ${\bf B}^2-{\bf E}^2$ can be expressed as:
\begin{equation*}
	({\bf B}^2 - {\bf E}^2)({\bf r}, t) = (b^2 - e^2)({\bf r})e^{-i\omega t}\;,
\end{equation*} 
where the term $(b^2 - e^2)$ depends only on ${\bf r}$ and doesn't depend on time $t$.
Then the solution to the equation \eqref{psi_eq_B2_E2} can be expressed as a delayed potential \cite{landau}:
\begin{equation} \label{Psi_sol_quadr}
	\Psi({\bf r}, t) = -\frac{a_1{\cal K}}{2\pi a_0}\int d^3 r' \frac{1}{|{\bf r} - {\bf r}'|}e^{-i(\omega t - k|{\bf r} - {\bf r}'|)}\big(b^2 - e^2\big)({\bf r}')\;.
\end{equation}
One can deduce from the equation \eqref{Psi_sol_quadr} that the dilaton field $\Psi$ is a superposition of spherical dilaton waves generated by the term $b^2 - e^2$, serving as a source density for the dilaton field. Furthermore, the source density of the dilaton field $(b^2 - e^2)$ can be determined in all of space.

\section{Basic equation} \label{sec:eq}     
                      
Suppose that at the origin of the Cartesian coordinates, there is a rotating neutron star (pulsar or magnetar) with a magnetic dipole moment $\bf m$. This star has a gravitational field. Since the distribution of matter in neutron stars is close to spherically symmetrical, one can consider the Schwarzschild solution as the metric tensor of pseudo-Riemannian space. It is convenient to write down this solution in isotropic coordinates. In these coordinates, the non-zero components of the metric tensor take the form \cite{landau}:
\begin{equation} \label{Scw_metric}
g_{00}={(4r-r_g)^2\over (4r+r_g)^2},\ \ \ \ \ \
g_{xx}=g_{yy}=g_{zz}=-(1+{r_g\over 4r})^4,
\end{equation}
where $r_g$ is the Schwarzschild radius and the following notation is introduced:
$r=\sqrt{x^2+y^2+z^2}$. Since the mass of neutron stars is less than 2 times the mass of the sun \cite{NS_mass}, a rough estimate can be made for $r_g$ of about 6 km.

The expression \eqref{Scw_metric} indicates that the gravitational field's influence on the dilaton formation process is restricted to a sphere with a radius of, say, about $10^6 r_g$. A region of space with a linear size of roughly 100 pc is where dilaton generation in the galactic magnetic field takes place. 
Accordingly, the gravitational fields of magnetars and pulsars have a minor effect on the radiation of magnetic dipole waves, changing their amplitudes by only 1 percent or less. Because of this, the influence of the gravitational field on the dilaton formation process can be ignored in this study. 

The magnetic vector potential ${\bf A}$ of an electromagnetic field in Minkowski space is expressed by the following equation, as is well known \cite{landau}:
\begin{equation} \label{eq_A}
	\big(\bigtriangleup - \frac{1}{c^2}\frac{\partial^2}{\partial t^2} \big)\ {\bf A}=
-{4\pi\over c}{\bf j}({\bf r},t),
\end{equation}
where ${\bf j}$ is a current density inside a neutron star and $c$ is the speed of light in a vacuum. 

Suppose that an electromagnetic field outside a star is analogous to a field created by rotating around axis $z$ magnetic dipole momentum $\bf m=\{m_x,m_y,m_z\}$ with frequency $\omega$:
\begin{equation} \label{m}
	{\bf m}(t)=|{\bf m}|\{\cos(\omega t)\sin\alpha, \ 
	\sin(\omega t)\sin\alpha,\ \cos\alpha \}\;.
\end{equation}
In the expression \eqref{m} $\alpha$ is an angle between magnetic dipole momentum ${\bf m}$ and axis $z$.
In this case, a neutron star radiates electromagnetic waves in magnetic dipole approximation.
One can write magnetic vector potential ${\bf A}$ in the following way:  
\begin{equation} \label{A}
\begin{split} 
	{\bf A}_w({\bf r}, \tau  ) = \frac{(\dot{{\bf m}}(\tau)\times {\bf r} )}{cr^2} 
+ \frac{({\bf m}(\tau) \times {\bf r})}{r^3}	\;.
\end{split}
\end{equation}
The dot above vector ${\bf m}$ represents the derivative on relative time $\tau = t - r/c$.

According to the Deutch solution \cite{Deutch_sol}, it is important to mention that a rotating neutron star can possess a quadrupole in addition to the dipole component \cite{ELD_NS}.
Nevertheless, in the distant region $r$ is much greater than $c/\omega$, the main component is the dipole.
Because dilatons are generated at cosmological distances approximately equal to $L_{\mathrm{coh}}$, the focus will be on the dipole component.

Electric strength and magnetic induction corresponding to magnetic vector potential ${\bf A}_w$ \eqref{A} can be expressed as (see also \cite{Ew_and_Bw}):
\begin{equation} \label{E_w}
{\bf E}_w({\bf r},\tau)=-\frac{1}{c}\frac{\partial {\bf A}_w({\bf r}, \tau)}{\partial \tau} =  \frac{({\bf r}\times\dot{\bf m}(\tau))}{ c r^3}+
\frac{({\bf r}\times\ddot{\bf m}(\tau))}{ c^2 r^2} 
\end{equation}
\begin{equation}\label{B_w}
\begin{split} 
{\bf B}_w({\bf r},\tau) = \mathrm{rot} {\bf A}_w
={3({\bf m}(\tau)\cdot {\bf r}){\bf r}
	-r^2{\bf m}(\tau)\over  r^5} \\
-{{ \dot{\bf m}}(\tau)\over c r^2}
+{3({ \dot{\bf m}}(\tau)\cdot {\bf r}){\bf r}\over c r^4}
+{(\ddot{\bf m}(\tau)\cdot {\bf r}){\bf r}-r^2\ddot{\bf m}(\tau)
	\over c^2 r^3}.
\end{split}
\end{equation}

Dilaton generation by electromagnetic fields ${\bf E}_w$ and ${\bf B}_w$ of rotating magnetic dipole momentum of pulsar or magnetar in the expressions \eqref{E_w}, \eqref{B_w} occurs at frequencies $\omega$ and $2\omega$ \cite{dil_from_pulsar}. Interesting to note that the arion generation by the same electromagnetic fields has the same spectrum \cite{arion_from_pulsar}.

Assume that there is a constant magnetic field ${\bf B_0}$ in the considered area of space.
\begin{equation} \label{B0}
	{\bf B_0} = \{B_{0x}, B_{0y}, B_{0z}\} \;.
\end{equation}

Such a magnetic field can be galactic and intergalactic fields, which can be considered constant at distances smaller than the coherence length $L_{\mathrm{coh}}$. The strength of the galactic magnetic field can be evaluated as $B_0 \sim 10^{-5}$ Gauss \cite{gltc_mag_field}.

In considered configuration of electromagnetic fields, the invariant \eqref{F2_inv} is equal:
\begin{equation} \label{2_inv}
	F_{nm}F^{nm}=
	2\big[{\bf B}_w^2-{\bf E}_w^2\big]+4({\bf B}_0\cdot {\bf B}_w)  + {\bf B}_0^2\;.
\end{equation}
 The term ${\bf B}_0^2$ at the end of equation \eqref{2_inv} does not change with time and will be disregarded in this study. 
Despite the enormous values of the magnetic field near the surface of a neutron star $B_{S}\sim 10^{8} $ - $10^{13}$ Gauss \cite{pulsar-catalog}, the term ${\bf B}_w^2-{\bf E}_w^2$ dominates at distances $r \ll c/\omega$ and decreases at $r \to \infty$ proportionally to $r^{-4}$. The second term, proportional to $({\bf B}_0\cdot {\bf B}_w)$, decreases at infinity as $r^{-1}$. For this reason, as we will see below, the second term makes a significant contribution to the intensity of dilaton generation, comparable to the intensity of dilaton generation from the term ${\bf B}_w^2-{\bf E}_w^2$.

\section{Solution of the dilaton field equation}  \label{sec:solution}        
Substituting the expression \eqref{2_inv} into \eqref{psi_equation_<<1} and excluding the terms ${\bf B}_w^2-{\bf E}_w^2$ considered in \cite{dil_from_pulsar} as well as the stationary term $B_0^2$, one can derive the following equation:
\begin{equation} \label{dilaton_eq}
(\Delta - {1\over c^2}{\partial^2 \over \partial t^2}) \Psi={4a_1{\cal K}\over a_0}({\bf B}_0 \cdot {\bf B}_w)\;.
\end{equation}
The equation \eqref{dilaton_eq} describes the dilaton generation during  propagation magnetic dipole radiation \eqref{E_w},\eqref{B_w} in the constant magnetic field \eqref{B0}.
Similar to equation \eqref {eq_A}, the term propotional to $({\bf B}_0 \cdot {\bf B}_w)$ in the right part of the equation \eqref{dilaton_eq} represents the source density of the dilaton field.

Considering the expression in \eqref{A}, \eqref{B_w} can be reformulated as follows:

\begin{equation} \label{B_w_rot}
	{\bf B_w}({\bf r}, t) = \mathrm{rot}\; \mathrm{rot} \frac{{\bf m}(\tau)}{r}\;.  
\end{equation} 

By substituting equation \eqref{B_w_rot} into equation \eqref{dilaton_eq}, it is possible to obtain:
\begin{equation} \label{dil_eq_in_glctc}
	(\Delta - {1\over c^2}{\partial^2 \over \partial t^2}) \Psi = \frac{4 {\cal K} a_1}{a_0}({\bf B_0}\cdot \mathrm{rot}\;\mathrm{rot}\frac{{\bf m}(\tau)}{r})\;.
\end{equation}
Finding a solution to the equation \eqref{dil_eq_in_glctc} can be easily done in the following manner:
\begin{subequations}
\begin{equation}\label{psi_and_F}
	\Psi = \frac{4 {\cal K} a_1}{a_0}\Big({\bf B_0}\cdot \mathrm{rot}\;\mathrm{rot}\; {\bf F}({\bf r}, t) \Big)\;. 
\end{equation}

Therefore, the equation for the vector-function ${\bf F}$ will be:
\begin{equation}\label{eq_for_F}
	(\Delta - {1\over c^2}{\partial^2 \over \partial t^2}){\bf F}({\bf r}, t) = \frac{{\bf m}(\tau)}{r} \;. 
\end{equation}

Place expression \eqref{m} on the right side of equation \eqref{eq_for_F} and write it in complex form:
\begin{equation} \label{eq_for_F_v2}
	(\Delta - {1\over c^2}{\partial^2 \over \partial t^2}){\bf F}({\bf r}, t) = |m|\sin\alpha({\bf e_x} + i{\bf e_y})\frac{1}{r}e^{-i(\omega t - kr)} \;,
\end{equation}
where $k = \omega /c$.

The solution to equation \eqref{eq_for_F_v2} can be found in the following form:
\begin{equation} \label{F_and_Q}
	{\bf F}({\bf r}, t) = |m|\sin\alpha({\bf e_x} + i{\bf e_y}) Q({\bf r})\;. 
\end{equation}

In this situation, the following equation for the function $Q$ can be derived:
\begin{equation} \label{eq_for_Q}
	(\Delta - {1\over c^2}{\partial^2 \over \partial t^2})Q({\bf r}, t) = 
\frac{1}{r}e^{-i(\omega t - kr)}\;. 
\end{equation}

By solving equation \eqref{eq_for_Q}, determine the unknown function $Q$:
\begin{equation} \label{Q_sol}
	Q({\bf r}, t) = \frac{1}{2ik}e^{-i(\omega t - kr)}\;.
\end{equation}
\end{subequations}

Finally, the exact solution of equation \eqref{dil_eq_in_glctc} is given by:
\begin{equation} \label{psi_sol}
	\begin{split}
	\Psi = \frac{2{\cal K}a_1 m   \sin\alpha}{a_0}\Big\{k(B_{0x} - ({\bf B_0 \cdot N})N_x)\sin(kr -\omega t) \\ 
	+k(B_{0y} - ({\bf B_0 \cdot N})N_y)\cos(kr-\omega t)  \\ 
	-\frac{1}{r}(B_{0x} + ({\bf B_0\cdot N})N_x)\cos(kr-\omega t) \\
	+ \frac{1}{r}(B_{0y} + ({\bf B_0\cdot N})N_y)\sin(kr-\omega t)	
	\Big\} \;.
	\end{split}
\end{equation}
In the expression \eqref{psi_sol}, the following notation is introduced: ${\bf N} = \frac{{\bf r}}{r} = \{\sin\theta \cos\phi, \sin\theta \sin\phi, \cos\theta\}$. It is worth noting that as $r \to \infty$, the dilaton field $\Psi$ becomes an oscillating function and does not approach 0.

\section{Angular distribution of the dilaton generation} \label{sec:I}

By the definition \cite{landau}, the amount of dilaton energy $dI$ emitted by the source per unit of time through a solid angle $d\Omega$ is given by the formula:
\begin{equation*}
	\frac{dI}{d\Omega} = \lim_{r\to \infty} r({\bf W \cdot r})\;,
\end{equation*}
where ${\bf W}$ is the energy flux density vector.
The vector ${\bf W}$ is associated with components of the stress-energy tensor $T^{ik}$ in the following way: $W^{\alpha} = T^{0\alpha}$. 
The stress-energy tensor of the free dilaton field has the form:
\begin{equation*}
	T^{ik} = 2a_0g^{in}g^{km}\big\{\frac{\partial \Psi}{\partial x^n}\frac{\partial \Psi}{\partial x^m} - \frac{1}{2}g_{mn}g^{sq}\frac{\partial \psi}{\partial x^s}\frac{\partial \psi}{\partial x^q}\big\}\;.
\end{equation*}
The angular distribution of dilaton generation can be calculated in the same way as in \cite{dil_q_plane_wave}:
\begin{equation*} 
{dI\over d\Omega}=-2a_0 r
({\bf r \cdot \nabla} \Psi){\partial \Psi\over\partial t}\;.
\end{equation*}

Thus, the angular distribution of the dilaton generation during the propagation in vacuum of a magnetic dipole wave from a pulsar in a constant magnetic field, averaged over the period of the electromagnetic wave, is equal:
\begin{equation} \label{dI_dOmega_dil}
	\overline{\frac{dI}{d\Omega}} = \frac{4{\cal K}^2 a_1^2 m^2 k\omega \sin^2\alpha}{a_0} \Big\{k^2r^2\frac{di_1}{d\Omega} + 4kr \frac{di_2}{d\Omega} + 2\frac{di_3}{d\Omega}\Big\}\;,
\end{equation}
where the following notations are introduced:
\begin{subequations}
\begin{equation} \label{i1}
	\frac{di_1}{d\Omega}= B_{0x}^2+B_{0y}^2 - 2({\bf B_0}\cdot{\bf N})(B_{0x}N_x + B_{0y}N_y) + ({\bf B_0}\cdot{\bf N})^2(N_x^2 +N_y^2)\;, 
\end{equation}
\begin{equation}\label{i2}
	\frac{di_2}{d\Omega} = ({\bf B_0}\cdot{\bf N})(B_{0x}N_y - B_{0y}N_x) \;, 
\end{equation}
\begin{equation}\label{i3}
	\frac{di_3}{d\Omega} = B_{0x}^2 + B_{0y}^2 + ({\bf B_0}\cdot{\bf N})(B_{0x}N_x+ B_{0y}N_y)\;. 
\end{equation}
\end{subequations}

The term \eqref{i1} can be viewed as the dominant factor in the angular distribution of dilaton production \eqref{dI_dOmega_dil} as $r \to \infty$:
\begin{equation} \label{dI1_dOmega_dil}
	\begin{split}
	\overline{\frac{dI}{d\Omega}} = \frac{4{\cal K}^2 a_1^2 m^2 k^3\omega r^2 \sin^2\alpha}{a_0}\Big[B_{0x}^2+B_{0y}^2 - 2({\bf B_0}\cdot{\bf N}) \\ \times(B_{0x}N_x + B_{0y}N_y)  + ({\bf B_0}\cdot{\bf N})^2(N_x^2 +N_y^2) \Big]\;. 
	\end{split}
\end{equation}

Next, we will examine the maximum and minimum values of the function $\overline{dI}/d\Omega$ \eqref{dI1_dOmega_dil} with respect to the variables $\theta$ and $\phi$. To do this, let $\theta_{B_0}$ and $\phi_{B_0}$ be the azimuth and polar angles, respectively, along which the magnetic field ${\bf B_0}$ \eqref{B0} is oriented:
\begin{equation*}
	{\bf B_0} = B_0\{\sin\theta_{B_0}\cos\phi_{B_0},\;\; \sin\theta_{B_0}\sin\phi_{B_0},\;\; \cos\theta_{B_0}\};.
\end{equation*}
For definiteness, let $\theta_{B_0} \le \frac{\pi}{2}$ and $\phi \le \pi$.

The expression \eqref{dI1_dOmega_dil} becomes 0 for the following values $\theta$ and $\phi$:
\begin{equation*}
	\begin{split}
	 \theta = \theta_{B_0}, \; \phi = \phi_{B_0} \;, \\
	\theta = \pi - \theta_{B_0}, \; \phi = \phi_{B_0} +\pi, \\
	\theta = \frac{\pi}{2}, \;  \phi = \phi_{B_0} \;, \\
	\theta = \frac{\pi}{2}, \;   \phi = \phi_{B_0} +\pi
	\end{split}
\end{equation*}
and has a maximum value at:
\begin{equation*}
	\begin{split}
	\theta = \frac{\theta_{B_0}}{2} + \frac{3\pi}{4} \;, \phi = \phi_{B_0}\;,\\
	\theta = \frac{\pi}{4} - \frac{\theta_{B_0}}{2} \;, \phi = \phi_{B_0} +\pi\;.
	\end{split}
\end{equation*}

By integrating the expression \eqref{dI1_dOmega_dil} over the angles $\theta$ and $\phi$, one can obtain total intensities of dilaton radiation in all directions:
\begin{equation} \label{I_dil}
	\begin{split}
	\overline{I} = \frac{16\pi{\cal K}^2 c a_1^2 m^2 k^4 \sin^2\alpha}{15a_0}r^2\Big\{9(B_{0x}^2+B_{0y}^2) + 2 B_{0z}^2\Big\}\;
	\end{split}
\end{equation}

It is important to note that, in astrophysical conditions, the galactic magnetic field is well characterized by \eqref{B0} only for distances $r$ smaller than the galactic magnetic field's coherence length $L_{\mathrm{coh}}$.  Thus, the expressions  \eqref{dI_dOmega_dil}, \eqref{dI1_dOmega_dil}, \eqref{I_dil} are applicable only for $r < L_{\mathrm{coh}} \sim 100$ pc.

According to \cite{dil_from_pulsar}, the intensity $\bar{I}_{\mathrm{PSR}}$ of dilaton generation by electromagnetic field \eqref{E_w}, \eqref{B_w} of rotating magnetic dipole momentum of a pulsar (see the first term of \eqref{2_inv}) can be evaluated as follows:
\begin{equation*}
	\bar{I}_{\mathrm{PSR}} \sim \frac{c{\cal K}^2a_1^2 B_{\mathrm{S}}^4 k^6 R_{S}^{10}}{a_0} \;,
\end{equation*} 
where $R_S$ is a radius of a neutron star, $B_S$ is a magnetic field on the neutron star surface.
Taking into account that magnetic dipole momentum can be calculated as $m = B_{S}R_S^3$, one can get an estimation:
\begin{equation*}
	\frac{\bar{I}}{\bar{I}_{\mathrm{PSR}}} \sim \frac{B_0^2r^2}{B_{S}^2 k^2 R_S^4}\;.
\end{equation*}

$R_S \sim 10^6$ cm is the typical radius of a neutron star \cite{NS_radii}. 
The ratio $\bar{I}/\bar{I}_{\mathrm{PSR}}$  can be $\gg 1$ when taking into account a millisecond pulsar with a magnetic field $B_S \sim 10^9$ Gauss that is situated at a distance from Earth on the order of the coherence length of the galactic magnetic field $L_{\mathrm{coh}} \sim 100$ pc. As an illustration, consider PSR J0437+4715. $R_S \sim 10^6$ cm, $B_S \sim 5\times10^8$ Gauss, and a period of 5 ms are reported by \cite{pulsar-catalog}.

As one can see, the obtained total intensity \eqref{I_dil} of the dilaton generation is proportional to $r^2$ and correspondingly tends to $\infty$ at $r \to \infty$. It is important to know that the expression \eqref{I_dil} is based on the idea that magnetic dipole radiation of a pulsar propagates in a vacuum but not in the interstellar medium. The properties of the interstellar medium, especially its reflective index, as we will see below, modify the behavior of the total intensity \eqref{I_dil} at $r \to \infty$.

\section{Dilaton generation during propagation magnetic dipole waves in a constant magnetic field in the interstellar medium} \label{sec:I_n>1}
As it is known, the interstellar medium consists of atoms, the concentration of which can vary from $0.1$ to $100$ atoms per cubic centimeter in different regions of space \cite{interstel-medium}. 
To simplify calculations, the interstellar medium will be assumed to be a rarefied neutral gas with a low density distributed uniformly. In this case the medium has a constant reflective index $n = 1+\chi$, and $\chi$ can be evaluated for the frequency $\omega$ of magnetic dipole rotation as $\sim 10^{-15}$.

The equation for magnetic vector potential ${\bf \tilde{A}}_w$ \eqref{eq_A} in the medium with reflective index $n$ has the following form:
	\begin{equation} \label{eq_for_tilde_A}
	\big(\bigtriangleup - \frac{n^2}{c^2}\frac{\partial^2}{\partial t^2} \big)\ {\bf \tilde{A}}_w({\bf r}, t)=-{4\pi\over c}{\bf j}({\bf r},t)\;.
\end{equation}

The equation \eqref{eq_for_tilde_A} indicates that electromagnetic waves have a speed $c/n$, which is lower than the speed of light in a vacuum.
The magnetic vector potential $\tilde{{\bf A}}_w$ generated by a rotating magnetic dipole moment ${\bf m}$ \eqref{m} in the medium with reflective index $n$ will take the form:
\begin{equation} \label{tilde_A}
	{\bf \tilde{A}}_w({\bf r}, t  ) = \frac{n(\dot{{\bf m}}(\tilde{\tau})\times {\bf r} )}{cr^2} + \frac{({\bf m}(\tilde{\tau}) \times {\bf r})}{r^3}	\;.
\end{equation}
In the expression \eqref{tilde_A} the magnetic dipole momentum ${\bf m}$ \eqref{m} depends on  $\tilde{\tau}$ 
\begin{equation*}
	\tilde{\tau} = t - \frac{nr}{c} \;,
\end{equation*}
 but not on $\tau$.

Thus, the equation \eqref{dil_eq_in_glctc} will take the form:
\begin{equation} \label{dil_eq_in_glctc_n>1}
	(\Delta - {1\over c^2}{\partial^2 \over \partial t^2}) \Psi = \frac{4 {\cal K} a_1}{a_0}({\bf B_0}\cdot \mathrm{rot}\;\mathrm{rot}\frac{{\bf m}(\tilde{\tau})}{r})\;.
\end{equation}
By performing analogous calculations (see \eqref{psi_and_F} - \eqref{F_and_Q}) one can obtain the following equation for function $Q$:
\begin{subequations}
\begin{equation} \label{eq_for_Q_n>1}
	(\Delta - {1\over c^2}{\partial^2 \over \partial t^2})Q({\bf r}, t) = \frac{1}{r}e^{-i(\omega t - nkr)}\;. 
\end{equation}

The particular solution of the equation \eqref{eq_for_Q_n>1} is found to be:
\begin{equation} \label{Q_p}
	Q_p = -\frac{1}{(n^2-1)k^2}\frac{1}{r}e^{-i(\omega t - nkr)}\;.
\end{equation}
The obtained particular solution \eqref{Q_p} tends to $\infty$ at $n \to 1$. 
But in the vacuum case ( $n=1$), the function $Q$ \eqref{Q_sol} has a finite value.
To eliminate this singularity, a solution of homogeneous equation $Q_0$ must be added.
Thus, the final expression for function $Q$ is found to be: 
\begin{equation} \label{Q_sol_n}
	Q = -\frac{1}{(n^2-1)k^2}\frac{1}{r}e^{-i\omega t}\big[e^{inkr}- e^{ikr}\big]\;. 
\end{equation}
\end{subequations}

One should note that the solution \eqref{Q_sol_n} at $n \to 1$ tends to the expression \eqref{Q_sol}.

Thus, the exact solution of the dilaton field equation \eqref{dil_eq_in_glctc_n>1} in the medium with reflective index $n$ in the complex form is found to be:
\begin{align}\label{psi_for_n>1}
	\Psi = -\mathrm{Re} \; \; \frac{4{\cal K}a_1 m\sin\alpha}{a_0(n^2-1)k^2} e^{-i\omega t}\Big\{
	\frac{k^2}{r}\big[B_{0x} + iB_{0y} \nonumber \\ -  ({\bf B_0\cdot N}) (N_{x}+iN_{y})\big]
	\big[n^2e^{inkr} - e^{ikr}\big] 
	+\frac{ik}{r^2}\big[ B_{0x}+iB_{0y}\nonumber \\- 3({\bf B_0 \cdot N})(N_x+iN_y)\big]
	\big[ne^{inkr} - e^{ikr}\big]
	-\frac{1}{r^3}\big[ B_{0x}+iB_{0y} \nonumber \\- 3({\bf B_0 \cdot N})(N_x +iN_y)\big]
	\big[e^{inkr}-e^{ikr}\big]
	\Big\}\;.
\end{align}

Unlike the vacuum case \eqref{psi_sol}, the dilaton field $\Psi$ in the expression \eqref{psi_for_n>1} teds to $0$ at $r \to \infty$ as $r^{-1}$  The dilaton field $\Psi$ in this instance is a superposition of dilaton waves moving at various speeds, which is equal to $c$ and $c/n$. As a result, depending on the phase difference, the generated wave may be stronger or weaker.

Since the interstellar medium's reflecting index is not much different from $1$, one can put $n = 1+\chi$, $\chi \ll 1$. Then the phase difference between the dilaton waves in superposition is equal to $\chi kr$.  The expression \eqref{psi_for_n>1} can be rewritten as:
\begin{align} \label{phase_diff}
	\Psi =- \mathrm{Re} \;\frac{2{\cal K} a_1 m \sin\alpha}{a_0\chi k^2}e^{-i(\omega t - kr)}\Big\{\frac{k^2}{r}\big[B_{0x}+iB_{0y}\nonumber \\-({\bf B_{0} \cdot N})(N_x+iN_y)\big] 
	\big[(1+2\chi)e^{i\chi kr}-1\big]+ \frac{ik}{r^2}\big[B_{0x}+iB_{0y} \nonumber \\-3({\bf B_0\cdot N})(N_x+iN_y)\big]
	\big[(1+\chi)e^{i\chi kr}-1\big]
	- \frac{1}{r^3}\big[B_{0x}+iB_{0y} \nonumber \\-3({\bf B_0\cdot N})(N_x+iN_y)\big] 
	\big[e^{i\chi kr} -1\big]
	\Big\} \;.
\end{align}

By extending the Taylor series and ignoring terms of order $\chi$ and higher, one can derive the expression \eqref{psi_sol} in case $\chi kr \ll 1$. If not, there would be oscillations in the function $\Psi$, and its minimums and maximums are determined by the phase difference $\chi kr$ of the dilaton waves in superposition.
At maximum and minimum values, $\chi kr$ is a multiple of $\pi$ and $2\pi$, respectively. 
 
For further investigation of the angular distribution of dilaton energy, one can consider
dilaton field $\Psi_w$ in the wave zone, which consists of terms inverse proportional to $r$:
\begin{align*}
	\Psi_w = -\frac{2{\cal K}a_1 m\sin\alpha}{a_0 \chi r}\Big\{
	\big[B_{0x} - ({\bf B_0\cdot N})N_x\big] \nonumber \\
	\times\big[(1+2\chi)
	\cos(kr-\omega t + \chi kr)  - \cos(kr-\omega t)\big]  
	\nonumber \\- \big[B_{0y} - ({\bf B_0\cdot N})N_y\big]  \big[(1+2\chi)\sin(kr - \omega t + \chi kr) \nonumber \\- \sin(kr-\omega t)\big]
	\Big\}\;.
\end{align*}

As a result, the average over the period of magnetic dipole wave angular distribution of the dilaton generation in propagation of magnetic dipole waves in a constant magnetic field is equal:
\begin{align} \label{dI_dw_dil_chi}
	\overline{\frac{dI}{d\Omega}} = \frac{8{\cal K}^2a_1^2k\omega m^2\sin^2\alpha}{a_0\chi^2}\Big[1 - \cos\big(\chi kr\big)\Big] \Big[B_{0x}^2+B_{0y}^2 \nonumber  \\
	- 2({\bf B_0\cdot N})(B_{0x}N_x + B_{0y}N_y) +({\bf B_0\cdot N})^2(N_x^2 + N_y^2)\Big] \;.
\end{align}

One should note, that at $\chi kr \to 0$ expression \eqref{dI_dw_dil_chi} tends to \eqref{dI1_dOmega_dil}.

Total intensity of such dilaton generation averaged over the period of magnetic dipole wave is equal:
\begin{align} \label{I_dil_chi}
	\bar{I} =  \frac{32\pi{\cal K}^2a_1^2k\omega m^2 \sin^2\alpha}{15a_0\chi^2}\Big[1 -\cos\big(\chi kr\big)\Big] \nonumber \\
	\times\big[9(B_{0x}^2 + B_{0y}^2) + 2B_{0z}^2\big]\;.
\end{align}

As one can see,  $\cos(\chi kr)$ in the expression \eqref{I_dil_chi} can be expanded in Taylor series at $\chi kr \ll 1$, and total intensity of dilaton generation increases proportionally $r^2$ as in the vacuum case (see \eqref{I_dil}). 
However, the total intensities will exhibit oscillating behavior at a distance $r_0$ ($\chi k r_0 \sim 1$).

Thus, in the interstellar medium, the obtained intensity of the dilaton generation during the propagation of magnetic dipole radiation from a pulsar in a galactic magnetic field has a finite value.
This is due to the fact that the dilaton field $\Psi$ \eqref{psi_for_n>1} is a combination of dilaton waves moving at speeds $c$ and $c/n$ with a phase shift $\chi kr$ that varies with distance.
Therefore, at distances $r\ll 1/(\chi k) $, dilaton waves can either strengthen or weaken each other based on the phase difference.

\section{Conclusion} \label{sec:conclusion}

According to Maxwell-dilaton theory, dilatons can be generated by electromagnetic fields with a nonzero electromagnetic field invariant ${\bf B}^2 - {\bf E}^2$. 
Since rotating neutron stars have very strong magnetic fields $B_S\sim 10^8$ - $10^{13}$ Gauss \cite {pulsar-catalog}, they can be promising sources of dilatons.
The dilaton generation by electromagnetic fields \eqref{E_w}, \eqref{B_w} of a rotating magnetic dipole momentum of a neutron star was considered in the study \cite{dil_from_pulsar}. It was shown that the intensity of dilaton generation is proportional to
\begin{equation*}
	\bar{I}_{\mathrm{PSR}} \sim \frac{{\cal K}^2a_1^2 B_{\mathrm{S}}^4 \omega^6 R_{S}^{10}}{c^5a_0} \;,
\end{equation*} 
where $B_S$ and $R_S$ are magnetic field on the pulsar surface and neutron star radius, respectively, and $\omega$ is neutron star rotation frequency.
Dilaton generation occurs at $\omega$ and $2\omega$ frequencies.

This study examines the dilaton generation in propagation of the magnetodipole radiation of a rotating neutron star \eqref{E_w}, \eqref{B_w} in a galactic magnetic field considered constant \eqref{B0} at a distance $r$ less than the coherence length of the galactic magnetic field $L_{\mathrm {coh}} \sim 100$ pc \cite{gltc_mag_field, gltc_legth_coh}.

In this study, the exact solution \eqref{psi_sol} of the dilaton field equation in approximation $\Psi \ll 1$ \eqref{dilaton_eq} is found, and the total intensity  of dilaton generation \eqref{I_dil} is obtained.
It has been demonstrated that intensity of dilaton generation \eqref{I_dil} is proportional to $\omega^4r^2 $:
\begin{align*} 
	\overline{I} \sim \frac{{\cal K}^2a_1^2B_S^2R_S^6}{c^3a_0}w^4B_0^2r^2\,
\end{align*}
where $B_0$ is a galactic magnetic field. 
The obtained results are applicable on distance $r<L_{\mathrm{coh} }\sim 100$ pc.
Dilaton generation occurs only at the neutron star's rotational frequency $\omega$.

Compare the acquired intensity $\overline{I}$ \eqref{I_dil} of dilaton generation to the intensity $\bar{I}_{\mathrm{PSR}}$ obtained in \cite{dil_from_pulsar}.  To achieve this goal, take into account the ratio $\bar{I}/\bar{I}_{\mathrm{PSR}}$:
\begin{equation*}
	\frac{\bar{I}}{\bar{I}_{\mathrm{PSR}}} \sim \frac{B_0^2c^2r^2}{B_{S}^2 \omega^2 R_S^4}
\end{equation*}
Despite the small value of the galactic magnetic field $B_0 \sim 10^{-5}$ Gauss \cite{gltc_mag_field}, for millisecond pulsars with a magnetic field on the surface $B_S\sim 10^9$ Gauss, the ratio $\bar{I}/\bar{I}_{\mathrm{PSR}}$ can be $\gg 1$.

As an illustration, let consider pulsar PSR J0437+4715. Accordingly to \cite{pulsar-catalog}, $R_S \sim 10^6$ Gauss, $B_S \sim 5 \times 10^8$ Gauss, and a period of 5 ms. In this instance
\begin{equation*}
	\frac{\bar{I}}{\bar{I}_{\mathrm{PSR}}} \sim \big(\frac{r}{1 \text {  pc}}\big)^2\;.
\end{equation*} 
Hence, the ratio of $\bar{I}/\bar{I}_{\mathrm{PSR}}$ is greater than 1 for PSR J0437+4715 at distances greater than 1 pc and reaches a value of $10^4$ at $r \sim 100$ pc.
This is the reason why the dilaton generation during the propagation of magnetic dipole radiation in a galactic magnetic field can be valuable for millisecond pulsars with a magnetic field strength $B_S \sim 10^9$ Gauss and located approximately $100$ pc away from Earth.

As mentioned, the dilaton generation intensity \eqref{I_dil} obtained in this study rises as distance $r$ increases, reaching an unlimited value as $r$ approaches infinity. Nevertheless, it is important to mention that in actuality, magnetic dipole radiation doesn't propagate in a vacuum but in an interstellar medium with a density of $0.1$ to $100$ atoms per cm$^3$ in various space regions \cite{interstel-medium}.
In this study, the interstellar medium is viewed as a rarefied neutral gas with a low density distributed uniformly. In this scenario, the medium has a constant refractive index $n = 1 + \chi $, $\chi \ll 1$.

As in the case of vacuum the exact solution \eqref{psi_for_n>1} to the dilaton field equation \eqref{dilaton_eq} was found for a medium with reflective index $n=1+\chi$. It has been demonstrated that in this situation the dilaton field $\Psi$ \eqref{psi_for_n>1} is a superposition of dilaton waves propagating at speeds $c$ and $c/n$. The dilaton waves can either strengthen or weaken each other when traveling long distances due to a large phase difference $\chi kr$ \eqref{phase_diff}. For this reason, the intensity of dilaton generation \eqref{I_dil_chi} stops growing at a distance of approximately $r\approx r_0 = \frac{1}{k\chi}$ and is replaced by oscillations:
\begin{align}
	\bar{I} \sim \frac{{\cal K}^2a_1^2B_S^2R_S^6}{ca_0\chi^2}\omega^2B_0^2\big[1 -\cos(\chi \frac{\omega r}{c})\big]\;.
\end{align}
By considering the interstellar medium as a rarefied neutral gas with a low density distributed uniformly, one can approximate $\chi\approx 10 ^ {-15} $ and, thus, $r_0\approx 10^3$ pc.

\section{Acknowledgements}
This study was conducted within the scientific program of the National Center for Physics and
Mathematics, section $ \#5$ $<<$Particle Physics and Cosmology$>>.$ Stage 2023-2025.

\end{document}